\documentclass{aa}

\usepackage{txfonts}
\usepackage{graphicx}
\usepackage{natbib}
\usepackage{lscape}
\bibpunct{(}{)}{;}{a}{}{,}

\begin{document}

\title{The miniJPAS survey: White dwarf science with 56 optical filters}

\author{C.~L\'opez-Sanjuan\inst{\ref{cefcaua}}
\and P.~-E.~Tremblay\inst{\ref{warwick}}
\and A.~Ederoclite\inst{\ref{cefca}}
\and H.~V\'azquez Rami\'o\inst{\ref{cefcaua}}
\and A.~J.~Cenarro\inst{\ref{cefcaua}}
\and A.~Mar\'{\i}n-Franch\inst{\ref{cefcaua}}
\and J.~Varela\inst{\ref{cefcaua}}
\and S.~Akras\inst{\ref{IAASR}}
\and M.~A.~Guerrero\inst{\ref{IAA}}
\and F.~M.~Jim\'enez-Esteban\inst{\ref{INTA},\ref{VO}}
\and R.~Lopes de Oliveira\inst{\ref{UFS},\ref{ON}}
\and A.~L.~Chies-Santos\inst{\ref{UFRGS},\ref{SAO}}
\and J.~A.~Fern\'andez-Ontiveros\inst{\ref{cefca}}
\and R.~Abramo\inst{\ref{IFUSP}}
\and J.~Alcaniz\inst{\ref{ON}}
\and N.~Ben{\'i}tez\inst{\ref{IAA}}
\and S.~Bonoli\inst{\ref{DIPC},\ref{ikerbasque},\ref{cefca}}
\and S.~Carneiro\inst{\ref{IFUFB}}
\and D.~Crist\'obal-Hornillos\inst{\ref{cefca}}
\and R.~A.~Dupke\inst{\ref{ON},\ref{MU},\ref{Alabama}}
\and C.~Mendes de Oliveira\inst{\ref{USP}}
\and M.~Moles\inst{\ref{cefca}}
\and L.~Sodr\'e Jr.\inst{\ref{USP}}
\and K.~Taylor\inst{\ref{Instruments4}}
}

\institute{Centro de Estudios de F\'{\i}sica del Cosmos de Arag\'on (CEFCA), Unidad Asociada al CSIC, Plaza San Juan 1, 44001 Teruel, Spain\\\email{clsj@cefca.es}\label{cefcaua}   
        \and
        Department of Physics, University of Warwick, Coventry, CV4 7AL, UK\label{warwick} 
        \and
        Centro de Estudios de F\'{\i}sica del Cosmos de Arag\'on (CEFCA), Plaza San Juan 1, 44001 Teruel, Spain\label{cefca}
        \and
        Institute for Astronomy, Astrophysics, Space Applications and Remote Sensing, National Observatory of Athens, Penteli GR 15236, Greece\label{IAASR}
        \and
        Instituto de Astrof\'{\i}sica de Andaluc\'{\i}a, IAA-CSIC, Glorieta de la Astronom\'{\i}a s/n, 18008 Granada, Spain\label{IAA}
        \and
        Centro de Astrobiolog\'{\i}a (CSIC-INTA), ESAC Campus, Camino Bajo del Castillo s/n, 28692 Villanueva de la Ca\~nada, Spain\label{INTA}
        \and
        Spanish Virtual Observatory, 28692 Villanueva de la Ca\~nada, Spain\label{VO}
        \and
        Departamento de F\'{\i}sica, Universidade Federal de Sergipe, Av. Marechal Rondon S/N, 49100-000 S\~ao Crist\'ov\~ao, Brazil\label{UFS}
        \and
        Observat\'orio Nacional - MCTI (ON), Rua Gal. Jos\'e Cristino 77, S\~ao Crist\'ov\~ao, 20921-400 Rio de Janeiro, Brazil\label{ON}
        \and
        Instituto de F\'{\i}sica, Universidade Federal do Rio Grande do Sul (UFRGS), Av. Bento Gon\~calves, 9500, Porto Alegre, RS, Brazil\label{UFRGS}
        \and
        Shanghai Astronomical Observatory, Chinese Academy of Sciences, 80 Nandan Rd., Shanghai 200030, China\label{SAO}
        \and
        Instituto de F\'{\i}sica, Universidade de S\~ao Paulo, Rua do Mat\~ao 1371, CEP 05508-090, S\~ao Paulo, Brazil\label{IFUSP}
        \and
        Donostia International Physics Centre (DIPC), Paseo Manuel de Lardizabal 4, 20018 Donostia-San Sebastián, Spain\label{DIPC}
        \and
        IKERBASQUE, Basque Foundation for Science, 48013, Bilbao, Spain\label{ikerbasque}
        \and
        Instituto de F\'{\i}sica, Universidade Federal da Bahia, 40210-340, Salvador, BA, Brazil\label{IFUFB}
        \and
        University of Michigan, Department of Astronomy, 1085 South University Ave., Ann Arbor, MI 48109, USA\label{MU}
        \and
        University of Alabama, Department of Physics and Astronomy, Gallalee Hall, Tuscaloosa, AL 35401, USA\label{Alabama}
        \and
        Instituto de Astronomia, Geof\'{\i}sica e Ci\^encias Atmosf\'ericas, Universidade de S\~ao Paulo, 05508-090 S\~ao Paulo, Brazil\label{USP}
        \and
        Instruments4, 4121 Pembury Place, La Canada Flintridge, CA, 91011, USA\label{Instruments4}
}

\date{Received 18 March 2022 / Accepted XX XX XX}

\abstract
{}
{We analyze the white dwarf population in miniJPAS, the first square degree observed with $56$ medium-band, $145\ \AA$ in width optical filters by the Javalambre Physics of the accelerating Universe Astrophysical Survey (J-PAS), to provide a data-based forecast for the white dwarf science with low-resolution ($R\sim50$) photo-spectra.}
{We define the sample of the bluest point-like sources in miniJPAS with $r < 21.5$ mag, point-like probability larger than $0.5$, $(u-r) < 0.80$ mag, and $(g-i) < 0.25$ mag. This sample comprises $33$ sources with spectroscopic information, $11$ white dwarfs and $22$ QSOs. We estimate the effective temperature ($T_{\rm eff}$), the surface gravity, and the composition of the white dwarf population by a Bayesian fitting to the observed photo-spectra.}
{The miniJPAS data permit the classification of the observed white dwarfs into H-dominated and He-dominated with 99\% confidence, and the detection of calcium absorption and polluting metals down to $r \sim 21.5$ mag at least for sources with $7\,000 < T_{\rm eff} < 22\,000$ K, the temperature range covered by the white dwarfs in miniJPAS. The effective temperature is estimated with a $2$\% uncertainty, close to the $1$\% from spectroscopy. A precise estimation of the surface gravity depends on the available parallax information. In addition, the white dwarf population at $T_{\rm eff} > 7\,000$ K can be segregated from the bluest extragalactic QSOs, providing a clean sample based on optical photometry alone.}
{The J-PAS low-resolution photo-spectra provide precise and accurate effective temperatures and atmospheric compositions for white dwarfs, complementing the data from {\it Gaia}. J-PAS will also detect and characterize new white dwarfs beyond the {\it Gaia} magnitude limit, providing faint candidates for spectroscopic follow up.}

\keywords{white dwarfs -- surveys -- techniques:photometric -- methods:statistical}

\titlerunning{The miniJPAS survey. White dwarf science with 56 optical filters.}

\authorrunning{L\'opez-Sanjuan et al.}

\maketitle

\section{Introduction}
White dwarfs are the degenerate remnant of stars with masses lower than $8-10$ $M_{\odot}$ and the endpoint of the stellar evolution for more than 97\% of stars \citep[e.g.,][and references therein]{ibeling13,doherty15}. This makes them an essential tool to disentangle the star formation history of the Milky Way, the late phases of stellar evolution, and to understand the physics of condensed matter.

White dwarfs can be selected from the general stellar population thanks to their location in the Hertzsprung-Russell (H-R) diagram, typically being ten magnitudes fainter than main sequence stars of the same effective temperature. The pioneering analysis by \citet{russell1914} and \citet{hertzsprung1915} shows only one faint A-type star, $40$ Eri B, with the inclusion of Sirius B \citep{adams1915} and van Maanen 2 \citep{vanmaanen1917,vanmaanen1920} in the lower left hand corner of the H-R diagram by the end of that decade. The initial doubts about the high density derived for these objects were clarified during the next years thanks to the estimation of the gravitational redshift of Sirius B \citep{adams1925} and the proposal of electron degeneracy pressure as a counterbalance for the gravitational collapse caused by such condensed matter \citep{fowler1926}. Once established as an astrophysical object (see \citealt{holberg08}, for a detailed review), the systematic analysis of the white dwarf population begun.

The use of the H-R diagram to search for new white dwarfs was limited by the difficulties in the estimation of precise parallaxes, which are needed to obtain the luminosity of the objects. Because of this, the definition of photometric white dwarf catalogs was mainly based on the search of ultraviolet-excess objects, such as the Palomar-Green catalog \citep[PG,][]{PGS}, the Kiso survey \citep[KUV,][]{KUV1,KUV2}, or the Kitt Peak-Downes survey \citep[KPD,][]{KPD}; and using reduced proper motions (e.g., \citealt{NLTT,harris06,rowell11,GF15,munn17}). The spectroscopic follow up of these photometric catalogs revealed a diversity of white dwarf atmospheric compositions \citep{sion83,wesemael93}, with sources presenting hydrogen lines (DA type), \ion{He}{II} lines (DO), \ion{He}{I} lines (DB), metal lines (DZ), and featureless spectra (DC) among others. By the end of the XXth century, about $\sim3\,000$ white dwarfs with spectroscopic information and only $\sim 300$ with precise parallax measurements were discovered \citep{mccook99}. 

This difference further increased by one order of magnitude mainly thanks to the spectroscopy from the Sloan Digital Sky Survey \citep[SDSS,][]{sdss}. During almost $20$ years of observations, the different SDSS data releases increased above $20\,000$ the number of white dwarfs with spectroscopic information \citep{kleinman04,eisenstein06_dr4,kepler15,kepler16,kepler19}. At the same time, the absolute number of white dwarfs with precise parallaxes did not increase significantly \citep[e.g.,][]{leggett18}. 

The high-quality data from the {\it Gaia} mission \citep{gaia} turned around the situation. Thanks to {\it Gaia} parallaxes and photometry, the efficient use of the H-R diagram to define the white dwarf population became feasible, with more that $350\,000$ candidates discovered so far \citep{gentilefusillo19, GF21}. It also permits the definition of high-confidence volume-limited white dwarf samples \citep{ hollands18,jimenezesteban18,kilic20,mccleery20,gaia_edr3_nearby}.

Gathering spectral information of the {\it Gaia}-based samples is a key observational goal to push forward the white dwarf science in the forthcoming years. Current and planned multi-object spectroscopic surveys, such as the SDSS-V Milky Way mapper \citep{sdssv}, the Large Sky Area Multi-Object Fiber Spectroscopic Telescope (LAMOST, \citealt{lamost}), the William Herschel Telescope Enhanced Area Velocity Explorer (WEAVE, \citealt{weave}), the Dark Energy Spectroscopic Instrument (DESI, \citealt{desi_mws}), and the 4-metre Multi-Object Spectrograph Telescope (4MOST, \citealt{4most_wd}), are going to observe a hundred thousand spectra of white dwarfs. In addition, the low-resolution data from {\it Gaia} spectro-photometry ($R \sim 30 - 90$) will provide valuable information for the white dwarf population \citep{carrasco14_wd}.

The exploitation of the spectroscopic data above can be enhanced by the photometry from the Javalambre Physics of the accelerating Universe Astrophysical Survey\footnote{\url{http://www.j-pas.org/}} (J-PAS, \citealt{jpas}); comprising $56$ optical passbands with a full width at half maximum of ${\rm FWHM}\sim 145\,\AA$ which provide low-resolution photo-spectra ($R \sim 50$) over several thousand square degrees in the northern sky. The J-PAS already released its first square degree in the Extended Groth Strip area, named miniJPAS \citep{minijpas}, providing a unique data set to test the capabilities of low-resolution spectral information for white dwarf science. Furthermore, and given the comparable spectral resolution, the miniJPAS analysis also yields a data-based forecast for {\it Gaia} spectro-photometry. 

We aim to study the white dwarf population in miniJPAS with $56$ optical filters, forecasting its capabilities in the estimation of the effective temperature ($T_{\rm eff}$), the surface gravity ($\log {\rm g}$), the atmospheric composition (hydrogen versus helium dominated), the detection of polluting metals, and the discrimination of extragalactic quasi-stellar objects (QSOs) with similar broad-band colors. 

This paper is organised as follows. We present the miniJPAS data and the bluest point-like sample in Sect.~\ref{sec:data}. The Bayesian analysis of the sources is detailed in Sect.~\ref{sec:fitting}. The results and the comparison with spectroscopy are described in  Sect.~\ref{sec:wds}. The separation between white dwarfs and QSOs is explored in Sect.~\ref{sec:qso}. Finally, Sect.~\ref{sec:conclusion} is devoted to the discussion and the conclusions. Magnitudes are expressed in the AB system \citep{oke83}.

\section{Data and sample definition}\label{sec:data}

\subsection{MiniJPAS photometric data}
The miniJPAS was carried out at the Observatorio Astrof\'{\i}sico de Javalambre (OAJ, \citealt{oaj}), located at the Pico del Buitre in the Sierra de Javalambre, Teruel (Spain). The data were acquired with the 2.5m Javalambre Survey Telescope (JST250) and the JPAS-Pathfinder (JPF) camera, which was the first scientific instrument installed at the JST250 before the arrival of the JPCam \citep{jpcam,jpcam2}. The JPF instrument is a single $9200 \times 9200$ CCD located at the centre of the JST250 field of view (FoV) with a pixel scale of 0.23 arcsec pixel$^{-1}$, providing an effective FoV of 0.27 deg$^2$.

The J-PAS filter system comprises $54$ filters with a FWHM of $145\,\AA$ that are spaced every $\approx 100\,\AA$ from $3\,800\,\AA$ to $9\,100\,\AA$. They are complemented with two broader filters at the blue and red end of the optical range, with an effective wavelength of $3\,479\,\AA$ ($u$, ${\rm FWHM} = 509\,\AA$) and $9\,316\,\AA$ ($J1007$, ${\rm FWHM} = 635\,\AA$), respectively. This filter system was optimized for delivering a low-resolution ($R \sim 50$) photo-spectrum of each surveyed pixel. The technical description and characterization of the J-PAS filters are presented in \citet{marinfranch12}. Detailed information of the filters as well as their transmission curves can be found at the Spanish Virtual Observatory Filter Profile Service\footnote{\url{http://svo2.cab.inta-csic.es/theory/fps/index.php?mode=browse&gname=OAJ&gname2=JPAS}}. In addition, miniJPAS includes four SDSS-like broad-band filters, named $u_{\rm J}gri$.

The miniJPAS observations comprise four JPF pointings in the Extended Groth Strip area along a strip aligned at $45$ deg with respect to North at $({\rm RA},{\rm Dec}) = (215,+53)$ deg, amounting to a total area of $\sim1$ deg$^2$. The depth achieved (3 arcsec diameter aperture, 5$\sigma$) is fainter than $22$ mag for filters bluewards of $7\,500\ \AA$ and $\sim22$ mag for longer wavelengths. The images and catalogs are publicly available on the J-PAS website\footnote{\url{http://www.j-pas.org/datareleases/minijpas_public_data_release_pdr201912}}.

We used the photometric data obtained with \texttt{SExtractor} \citep{sextractor} in  the  so-called  dual mode. The $r$ filter was used as the detection band and, for the rest of the filters, the aperture defined in the reference $r$-band was used to extract the flux. The observed fluxes in a $3$ arcsec diameter aperture were stored in the vector $\vec{f} = \{ f_j \}$, and their errors in the vector $\sigma_{\vec{f}} = \{\sigma_j\}$, where the index $j$ runs the miniJPAS passbands. The error vector includes the uncertainties from photon counting and sky background.

Further details on the miniJPAS observations, data reduction, and photometric calibration can be found in \citet{minijpas}.

\subsection{Aperture correction of the $3$ arcsec photometry}\label{sec:3tototal}
The miniJPAS magnitudes measured in a $3$ arcsec aperture are not the total magnitudes of the sources and an aperture correction is needed. The aperture correction was defined as
\begin{equation}
    C^{\rm aper} = C^{\rm tot}_{6} + C^{6}_{3}.
\end{equation}
The first term is the correction from $6$ arcsec magnitudes to total magnitudes (i.e., the total flux of the star), that depends on the pointing and the filter. The second term corrects the $3$ arcsec photometry to the $6$ arcsec photometry and also depends on the position of the source in the CCD due to the variation of the point spread function along the FoV. The techniques and assumptions applied in the computation of these two aperture corrections are detailed in \citet{clsj22pda}.

\begin{figure}[t]
\centering
\resizebox{\hsize}{!}{\includegraphics{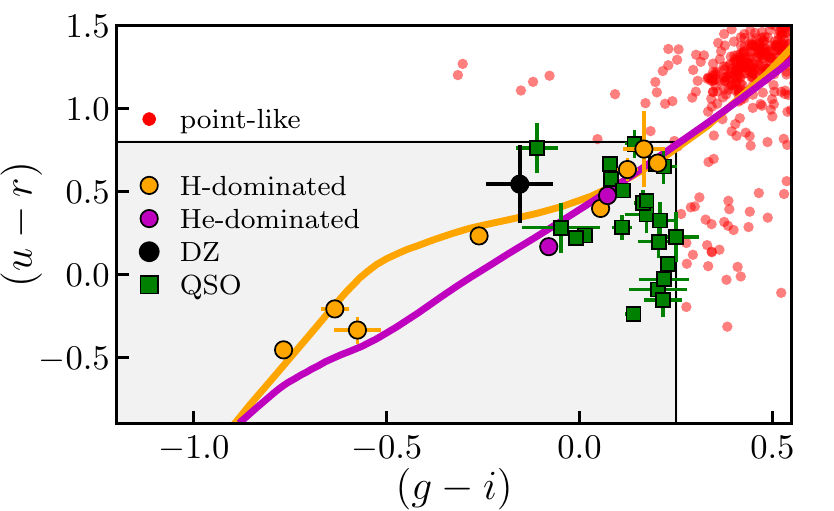}}
\caption{The $(u-r)$ versus $(g-i)$ color-color diagram in miniJPAS for point-like sources with $r < 21.5$ mag (red dots). The gray area defines the bluest point-like sources with $(u-r) < 0.80$ and $(g-i) < 0.25$. The $33$ sources in this area are classified from SDSS spectrum as $11$ white dwarfs ($8$ with H-dominated atmosphere, orange dots; $2$ with He-dominated atmosphere, purple dots; $1$ metal polluted DZ, black dot) and $22$ QSOs (green squares). The expected colors from theoretical cooling tracks for H-dominated and He-dominated atmospheres are shown with the orange and purple lines, respectively.}
\label{fig:bps}
\end{figure}

\begin{table*}
\caption{White dwarfs in the miniJPAS bluest point-like sample: photometry and {\it Gaia} information.}
\label{tab:wds}
\centering 
        \begin{tabular}{l c c c c c c c}
        \hline\hline\rule{0pt}{3ex} 
         Tile - Number   & RA    & Dec  & $r$  & $(g-i)$ & $(u-r)$ & $\varpi_{\rm EDR3}$ & Reference\\
         \rule{0pt}{2ex} 
                &   [deg]                & [deg]           &  [mag]  &  [mag]  & [mag] &  [mas] & \\
        \hline\rule{0pt}{2ex}
        \!$2241-1747$        &214.1778  &  52.2620   &       $19.32 \pm 0.01$   &  $-0.08 \pm 0.01$ &  $0.17\pm0.02$  & $4.36 \pm 0.34$ & 1\\  
        
        $2241-19527$         &214.5030  &  52.4108   &       $19.61\pm0.01$   & $-0.26\pm0.02$ & $0.23\pm0.03$ & $4.95 \pm 0.24$ & 1,2\\ 
        
        $2243-2625$          &214.3501  &  52.8745   &       $19.30\pm0.01$   & $0.05\pm0.01$ & $0.40\pm0.04$ & $5.37 \pm 0.19$ & 1,2\\  

        $2243-4859$          &214.7533  &  52.7319   &       $21.48\pm0.04$   & $-0.15\pm0.09$  & $0.54\pm0.24$ & $\cdots$ & $\cdots$ \\ 
        
        $2243-5175$          &214.9642  &  52.6212   &       $21.12\pm0.03$   & $-0.58\pm0.06$ & $-0.34\pm0.08$ & $-2.6 \pm 1.5$ & 1\\  
        
        $2406-5601$          &215.3567    & 53.0818   &      $19.00\pm0.01$   & $0.20\pm0.01$ & $0.67\pm0.03$ & $7.35 \pm 0.15$ & 1,2\\    
        
        $2406-9645$          &215.1359   & 53.2736   &       $19.33\pm0.01$   & $0.07\pm0.01$ & $0.47\pm0.03$ & $6.31 \pm 0.19$ & 1,2 \\  

        $2406-16326$         &215.7065   & 53.0915   &       $21.12\pm0.03$   & $0.17\pm0.06$  &  $0.76\pm0.23$ & $\cdots$ & $\cdots$  \\ 
        
        $2470-3588$          &213.4513    & 52.1572   &       $19.84\pm0.01$   & $-0.77\pm0.02$ & $-0.46\pm0.02$ & $1.49 \pm 0.24$ & 1,2 \\   
        
        $2470-13619$         &214.0558    &  52.1936   &       $20.46\pm0.01$   & $0.12\pm0.03$ & $0.63\pm0.07$ & $1.99 \pm 0.55$ & 1,2\\    
        
        $2470-15262$         &214.0462   &  52.1328   &       $20.55\pm0.02$   & $-0.63\pm0.04$ & $-0.21\pm0.04$ & $0.37 \pm 0.53$ & 1\\    
        
        \hline 
        \end{tabular}
    \tablebib{
    (1)~\citet{gaiaedr3}; (2)~Included in the \citet{GF21} catalog based on {\it Gaia} EDR3 with white dwarf probability $P_{\rm WD} > 0.75$.
    }
\end{table*}

\begin{table*}
\caption{White dwarfs in the miniJPAS bluest point-like sample: Atmospheric parameters from SDSS spectroscopy.}
\label{tab:wds_spec}
\centering 
        \begin{tabular}{l c c c c c c}
        \hline\hline\rule{0pt}{3ex} 
         Tile - Number   &   SDSS  & Type & Composition & $T_{\rm eff}^{\rm spec}$ & $\log {\rm g}^{\rm spec}$ & Reference\\
         \rule{0pt}{2ex}
                & Plate-MJD-Fiber & & & [K] & [dex] &  \\
        \hline\rule{0pt}{2ex}
        \!$2241-1747$        &  7028-56449-0220  & DA  & He  &  $5240 \pm 130$  &  $7.6 \pm 0.3$ & 2 \\
        
        $2241-19527$         &  7028-56449-0185  & DA & H   & $11\,010 \pm 100$  & $8.70 \pm 0.06$ & 2 \\
        
        $2243-2625$          &  7028-56449-0199  & DA  & H  & $7\,880 \pm 60$   & $7.85 \pm 0.11$  & 2 \\  

        $2243-4859$         &   7030-56448-0227  & DZ  & He  & $\cdots$ & $\cdots$ & $\cdots$ \\
        
        $2243-5175$          &  7028-56449-0102  & DA  & H   & $19\,300 \pm 700$   & $8.17 \pm 0.11$ & 1\\  
        
        $2406-5601$          &  7028-56449-0930  & DA  & H   & $7\,380 \pm 50$   & $7.78 \pm 0.11$  & 2\\ 
        
        $2406-9645$          &  6717-56397-0721  & DC & He  & $\cdots$ & $\cdots$ & $\cdots$ \\  
        
        $2406-16326$         &  7031-56449-0616  & DA & H  & $7\,400 \pm 300$   & $8 \pm 1$    & 1\\
        
        $2470-3588$          &  7030-56448-0453  & DA  & H & $22\,800 \pm 400$   & $7.89 \pm 0.06$  &  2\\   
        
        $2470-13619$         &  7030-56448-0380 & DA & H &   $8\,520 \pm 120$   &  $7.7 \pm 0.2$    &  2\\   
        
        $2470-15262$         &  7029-56455-0253	 & DA & H &  $17\,600 \pm 500$   & $7.76 \pm 0.09$   & 1\\   
        \hline 
        \end{tabular}
   \tablebib{
    (1)~\citet{kepler16}; (2) \citet{kepler19}.
    }
\end{table*}

\begin{table*}
\caption{White dwarfs in the miniJPAS bluest point-like sample: fitting results from miniJPAS photometry.}
\label{tab:wds_jpas}
\centering 
        \begin{tabular}{l c c c c c}
        \hline\hline\rule{0pt}{3ex} 
         Tile - Number   & $p_{\rm H}$ & $T_{\rm eff}$ & $\log {\rm g}$ & $\varpi$  & $\chi^2_{\rm WD}$\\
         \rule{0pt}{2ex} 
                & & [K] & [dex] & [mas] & \\
        \hline\rule{0pt}{2ex}
        \!$2241-1747$        &   $0.00$   & $8\,510 \pm 120$ & $7.69 \pm 0.14$ & $4.4 \pm 0.3$  &  $20.0$\tablefootmark{a} \\  
        
        $2241-19527$         &  $1.00$   & $11\,250 \pm 240$   & $8.66 \pm 0.06$ & $4.99 \pm 0.24$   &  $33.7$\tablefootmark{a}\\ 
        
        $2243-2625$          &   $1.00$   & $7\,850 \pm 100$   & $7.91 \pm 0.07$  & $5.40 \pm 0.19$   &  $18.4$\tablefootmark{a}\\  

        $2243-4859$      &   $0.88$   &  $8\,700 \pm 400$   & $7.0 \pm 0.7$ &  $1.16 \pm 0.23$    &  $81.3$\tablefootmark{a}\\
        
        $2243-4859$        &   $0.01$   &  $8\,800 \pm 400$   & $8.3 \pm 0.6$ &  $2.2 \pm 1.0$    &  $53.2$\tablefootmark{b}\\
        
        $2243-5175$          &   $1.00$   & $19\,900 \pm 1\,100$   & $7.8 \pm 0.5$ & $0.9 \pm 0.3$   &  $52.9$\tablefootmark{a}\\  
        
        $2406-5601$          &   $1.00$   &  $7\,160 \pm 80$   & $7.95 \pm 0.04$   & $7.35 \pm 0.15$   &  $13.2$\tablefootmark{a}\\    
        
        $2406-9645$          &   $0.00$   & $7\,510 \pm 90$   & $7.99 \pm 0.06$  &   $6.30 \pm 0.19$   &  $36.7$\tablefootmark{a}\\  
        
        $2406-16326$         &   $0.99$   & $7\,150 \pm 180$   & $8.8 \pm 0.7$ & $5.2\pm 3.4$    &  $48.3$\tablefootmark{a}\\ 
        
        $2470-3588$          &  $1.00$   & $22\,400 \pm 900$   & $7.87 \pm 0.22$  &  $1.45 \pm 0.20$   &  $47.8$\tablefootmark{a}\\   
        
        $2470-13619$         &  $1.00$   &   $8\,140 \pm 180$   &  $7.1 \pm 0.4$  & $1.99 \pm 0.25$   &  $44.9$\tablefootmark{a}\\    
        
        $2470-15262$         &   $1.00$  &  $18\,000 \pm 800$   & $7.6 \pm 0.3$   & $1.02 \pm 0.21$   &  $52.8$\tablefootmark{a}\\    
        \hline 
        \end{tabular}
    \tablefoot{
    \tablefoottext{a}{All the miniJPAS passbands were used in the fitting.}
    \tablefoottext{b}{Passbands $J0390$ and $J0400$ were removed in the analysis (Sect.~\ref{sec:dz}).}
    }
\end{table*}

\begin{table*}
\caption{Quasars in the miniJPAS bluest point-like sample.}
\label{tab:qso}
\centering 
        \begin{tabular}{l c c c c c c c c}
        \hline\hline\rule{0pt}{3ex} 
         Tile - Number   & RA    & Dec  & $r$  & $(g-i)$ & $(u-r)$ & $\varpi_{\rm EDR3}$ & $z_{\rm spec}$ & $\chi^2_{\rm WD}$\\
         \rule{0pt}{2ex} 
                &   [deg]                & [deg]           &  [mag] &  [mag] &  [mag] &   [mas]    &     &    \\
        \hline\rule{0pt}{2ex}
        \!$2241-3755$        &213.9637  & 52.4613          &       $19.73\pm0.01$   & $0.20\pm0.02$ & $0.66\pm0.04$ & $-0.20 \pm 0.28$  &     $2.581$  &  $314.0$ \\
        
        $2241-7683$          &214.3127  &        52.3869   &       $21.24\pm0.03$   & $0.20\pm0.07$ & $-0.09\pm0.08$ & $\cdots$  &   $ 	1.260$  &  $131.4$\\ 
        
        $2241-9344$          &214.4118  &        52.3925   &       $21.26\pm0.03$   & $0.17\pm0.06$ & $0.36\pm0.11$ & $\cdots$  &    $ 	2.159$  &  $472.2$\\

        $2241-14404$         &214.3972  &        52.6476   &       $20.24\pm0.01$   & $0.14\pm0.02$ & $-0.24\pm0.03$ & $0.17 \pm 0.44$  &     $1.961$  &  $520.8$  \\ 
        
        $2241-20770$         &214.5971  &        52.6679   &       $21.28\pm0.03$   & $0.21\pm0.05$ & $0.20\pm0.10$ & $\cdots$  &     $ 	1.766$  &  $284.1$\\  
        
        $2243-12132$         &215.0233  &        53.0102   &       $19.81\pm0.01$   & $0.23\pm0.02$ & $0.06\pm0.04$ & $-0.18 \pm 0.28$  &     $1.647$  &  $431.4$\\
        
        $2243-12352$         &214.9700  &        53.0345  &       $20.47\pm0.02$  & $0.11\pm0.03$ & $0.28\pm0.07$ & $1.76 \pm 0.74$  &    $1.902$  &  $265.3$\\ 
        
        $2243-12363$         &214.9946  &       53.0194   &       $20.83\pm0.02$   & $0.21\pm0.04$ & $0.32\pm0.11$ & $-1.03 \pm 0.72$  &     $1.728$  &  $200.7$\\
        
        $2406-853$          &215.2185   &       52.9396   &       $18.14\pm0.01$   & $0.01\pm0.01$ & $0.23\pm0.01$ & $0.09 \pm 0.08$  &     $0.676$  &  $430.4$\\
        
        $2406-1224$         &215.3250   &       52.8961   &       $20.39\pm0.01$   & $0.16\pm0.02$ & $0.43\pm0.08$ & $0.43 \pm 0.37$  &     $2.305$  &  $577.3$\\
        
        $2406-4342$         &215.0437   &       53.2066   &       $20.09\pm0.01$   & $0.14\pm0.02$ & $0.79\pm0.09$ & $-0.68 \pm 0.46$  &     $2.590$  &  $263.8$\\ 
        
        $2406-5133$        &215.3823    &       53.0450   &       $21.26\pm0.03$   & $-0.05\pm0.10$ & $0.28\pm0.15$ & $\cdots$  &             $0.957$  &  $145.0$\\
        
        $2406-8977$         &215.3053   &       53.2052   &       $21.21\pm0.03$   & $0.22\pm0.05$ & $-0.16\pm0.10$ & $\cdots$  &     $ 	1.953$  &  $204.5$\\
        
        $2406-11608$        &215.7752   &       53.2581   &       $18.44\pm0.01$   & $0.08\pm0.01$ & $0.67\pm0.02$ & $-0.15 \pm 0.13$  &   $2.462$  &  $481.1$\\

        $2406-14008$        &215.6897   &       53.2010   &       $21.34\pm0.03$   & $0.25\pm0.06$ & $0.26\pm0.15$ & $\cdots$  &     $1.671$  &  $99.6$   \\ 
        
        $2406-14869$        &215.4022   &       53.3373   &       $21.48\pm0.04$   & $0.22\pm0.06$ & $-0.03\pm0.14$ & $\cdots$  &     $2.019$  &  $160.9$   \\ 
        
        $2470-2363$         &213.7993   &        51.8822  &       $19.60\pm0.01$   & $0.08\pm0.01$ & $0.57\pm0.03$ & $-0.01 \pm 0.22$  &     $2.306$  &  $350.4$\\
        
        $2470-4230$         &213.4213   &        52.2056   &      $18.96\pm0.01$  & $0.17\pm0.01$ & $0.44\pm0.02$ & $-0.05 \pm 0.21$  &     $1.213$  &  $127.7$ \\
        
        $2470-4455$          &213.4495  &        52.2014   &       $21.16\pm0.03$   & $-0.11\pm0.05$ & $0.76\pm0.15$ & $\cdots$  &    $2.351$  &  $159.8$ \\
        
        $2470-7732$         &213.8912   &        52.0994   &       $18.15\pm0.01$   & $-0.01\pm0.01$ & $0.22\pm0.01$ & $-0.03 \pm 0.10$  &     $0.987$  &  $198.3$ \\
        
        $2470-9064$          &213.4318  &        52.3207   &       $20.49\pm 0.02$  & $0.22\pm0.03$ & $0.65\pm0.11$ & $-0.94 \pm 0.62$  &     $1.327$  &  $258.8$ \\
        
        $2470-13393$         &213.6948  &        52.4233   &       $19.47\pm0.01$  & $0.11\pm0.01$ & $0.51\pm0.03$ & $-0.48 \pm 0.26$  &     $1.583$  &  $313.7$ \\
        
        \hline 
        \end{tabular}
\end{table*}

\subsection{The sample of bluest point-like sources}
We aim to provide a forecast for the white dwarf science with the J-PAS 56 optical filters. Our working sample is composed of the bluest point-like sources (BPS) in the miniJPAS catalog, where mainly white dwarfs and extragalactic QSOs are expected. To define the BPS catalog, magnitudes from $3$ arcsec photometry corrected by aperture effects were used.  

The point-like sample was defined with apparent magnitude $r < 21.5$ mag to ensure a large enough ${\rm S/N}$ in the medium-band photometry. This magnitude selection translates to a median ${\rm S/N}$ per passband larger than $5$ in all the BPS. We also imposed a probability of being point-like of $p_{\rm point} > 0.5$. We obtained a total of $2\,684$ sources with these criteria.

Then, a colour selection was performed in the $(u-r)$ versus $(g-i)$ color diagram (Fig.~\ref{fig:bps}). We used these colors to ensure independent measurements and avoid correlated uncertainties. Several structures are apparent in this color-color plot. White dwarfs occupy the bluest corner in the plot, as illustrated with the H-dominated and He-dominated theoretical cooling tracks (see Sect.~\ref{sec:fitting}, for details about the assumed models). A sparsely populated sequence, corresponding to A-type and blue horizontal branch stars, is at $(g-i) \lesssim 0.3$ mag and $(u-r) \sim 1.2$ mag. The common F-type stars produce the overdensity at $(g-i) \sim 0.4$ mag and $(u-r) \sim 1.3$ mag. Finally, the QSO population is responsible for the data excess visible at $(g-i) \sim 0.3$ and $(u-r) \sim 0.3$ mag.

We defined the bluest sources in miniJPAS with $(u-r) < 0.80$ mag and $(g-i) < 0.25$ mag. These color selections ensure a complete sample for white dwarfs at $T_{\rm eff} \gtrsim 7\,000$ K, as expected from the theoretical cooling tracks described in Sect.~\ref{sec:fitting}, and minimise the contamination from main sequence stars and QSOs. The selection provided a total of $33$ sources in the surveyed area of one square degree, defining the BPS sample.

The next step was to gather all the available information about the BPS in the literature. We searched for information in the Montreal white dwarf database\footnote{\url{http://www.montrealwhitedwarfdatabase.org}} \citep{MWDD} and Simbad\footnote{\url{http://simbad.u-strasbg.fr/simbad}} \citep{simbad}. We also collected SDSS spectroscopy and {\it Gaia} EDR3 \citep{gaiaedr3} astrometry. We found that all the BPS have a SDSS spectrum, providing a spectral classification of the sources. The BPS sample was split into $11$ white dwarfs (Tables~\ref{tab:wds}, \ref{tab:wds_spec}, and \ref{tab:wds_jpas}) and $22$ QSOs (Table~\ref{tab:qso}).

We study the physical properties of the $11$ white dwarfs in Sect.~\ref{sec:wds}, and state the capabilities of miniJPAS photometry to disentangle between white dwarfs and QSOs in Sect.~\ref{sec:qso}.

\section{Bayesian estimation of white dwarf atmospheric parameters and composition}\label{sec:fitting}
The Bayesian methodology used to analyse the miniJPAS data was developed in \citet{clsj22pda} to study the white dwarf population in the Javalambre Photometric Local Universe Survey (J-PLUS, \citealt{cenarro19}), comprising $12$ optical filters. We adapted the method to deal with the $56$ medium bands in miniJPAS and included in the analysis the $g$ and $i$ broad bands. The $u_{\rm J}$ and $r$ broad bands were not used because they have been discarded from the final J-PAS observing strategy, that will only include $g$ and $i$. We provide in the following a summary of the fitting process, that is fully detailed in \citet{clsj22pda}.

We estimated the normalized probability density function (PDF) for each white dwarf in the sample,
\begin{equation}
    {\rm PDF}\,(t,\theta\,|\,\vec{f}, \vec{\sigma}_{\vec{f}}) \propto \mathcal{L}\,(\,\vec{f}\,|\,t,\theta,\sigma_{\vec{f}}) \times P\,(\varpi),
\end{equation}
where $t = \{ {\rm H}, {\rm He} \}$ are the explored H- and He-dominated atmosphere compositions, $\theta = \{T_{\rm eff}, \log g, \varpi \}$ are the parameters in the fitting (effective temperature, surface gravity, and parallax), $\mathcal{L}$ is the likelihood of the data for a given set of parameters and atmospheric composition, and $P$ is the prior probability imposed to the parallax.

The likelihood was defined as
\begin{equation}
    \mathcal{L}\,(\,\vec{f}\,|\,t,\theta,{\boldmath \sigma}_{\vec{f}}) = \prod_{j = 1}^{58} P_{\rm G}\,(f_j\,|\,f_{t,j}^{\rm mod}, \sigma_{j}),
\end{equation}
where the index $j$ runs over the $56$ medium bands and the $gi$ broad bands in miniJPAS, the function $P_{\rm G}$ defines a Gaussian distribution with median $\mu$ and dispersion $\sigma$, and the model flux was
\begin{equation}
    f_{t,j}^{\rm mod}\,(t,\theta)= \bigg( \frac{\varpi}{100} \bigg)^2\,F_{t,k}\,(T_{\rm eff},\log g)\,10^{0.4\,k_j\,E(B-V)}\,10^{0.4\,C^{\rm aper}_j},
\end{equation}
where  $k_j$ is the extinction coefficient of the filter, $E(B-V)$ the colour excess of the source, $C^{\rm aper}_j$ is the aperture correction needed to translate the observed $3$ arcsec fluxes to total fluxes (Sect.~\ref{sec:3tototal}), and $F_{t,k}$ is the theoretical absolute flux emitted by a white dwarf at 10 pc distance. The uncertainty in the photometric calibration ($\sigma_{\rm cal} = 0.04$ mag) was included in the error vector. 

The color excess was estimated by using the 3D reddening map from \citet{bayestar17}\footnote{We used the \texttt{Bayesta17} version of the map, available at \url{http://argonaut.skymaps.info}} at distance $d = \varpi^{-1}$. We note that this extinction correction was used in the photometric calibration of miniJPAS, so we also used it for consistency.

Pure-H models were assumed to describe H-dominated atmospheres ($t = {\rm H}$, \citealt{tremblay11,tremblay13}). Mixed models with H/He = $10^{-5}$ at $T_{\rm eff} > 6\,500$ K and pure-He models at $T_{\rm eff} < 6\,500$ K were used to define He-dominated atmospheres ($t = {\rm He}$, \citealt{cukanovaite18, cukanovaite19}). The mass-radius relation of \citet{fontaine01} for thin (He-atmospheres) and thick (H-atmospheres) hydrogen layers were used in the modelling. The justification of these choices and extended details about the assumed models can be found in \citet{bergeron19,GF20,GF21}; and \citet{mccleery20}.

\begin{figure*}[hp]
\centering
\resizebox{0.49\hsize}{!}{\includegraphics{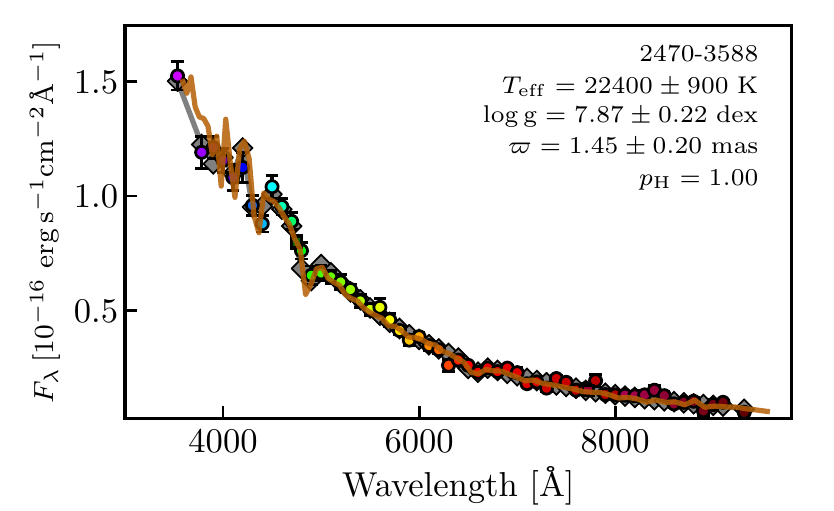}}
\resizebox{0.49\hsize}{!}{\includegraphics{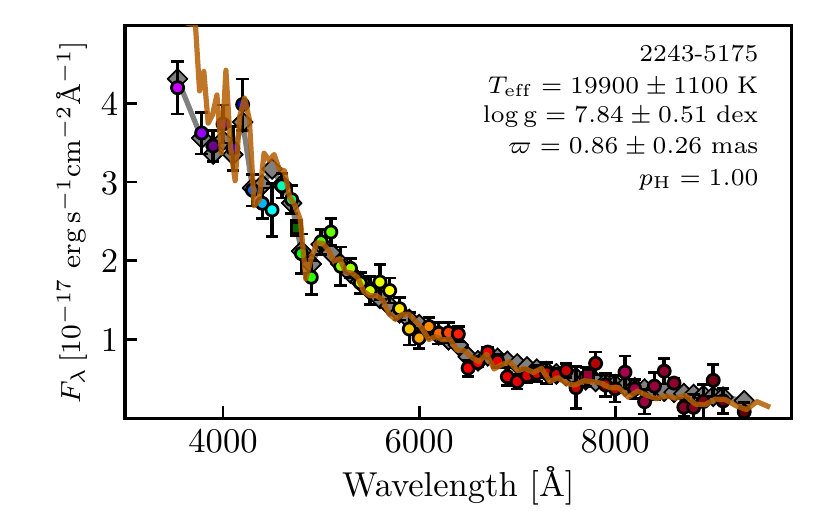}}\\
\resizebox{0.49\hsize}{!}{\includegraphics{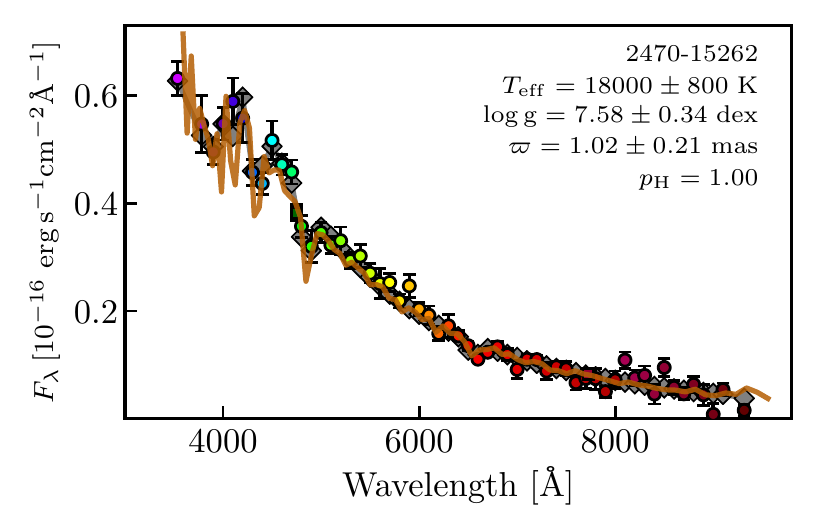}}
\resizebox{0.49\hsize}{!}{\includegraphics{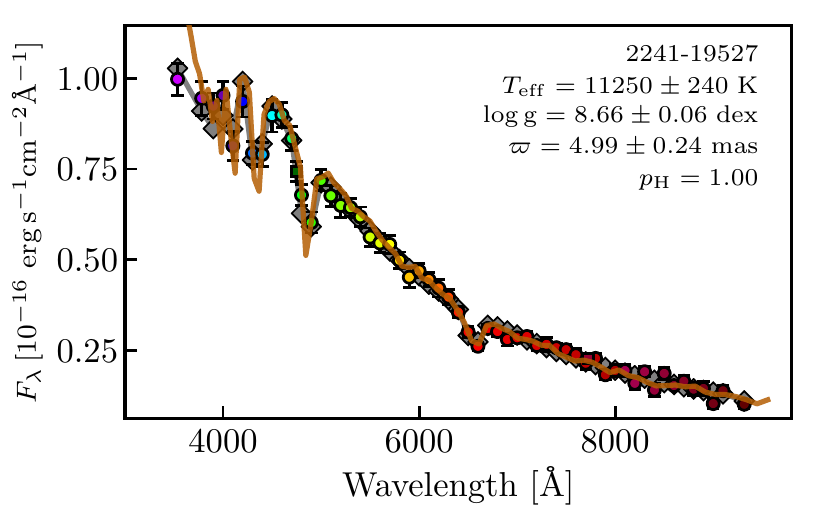}}\\
\resizebox{0.49\hsize}{!}{\includegraphics{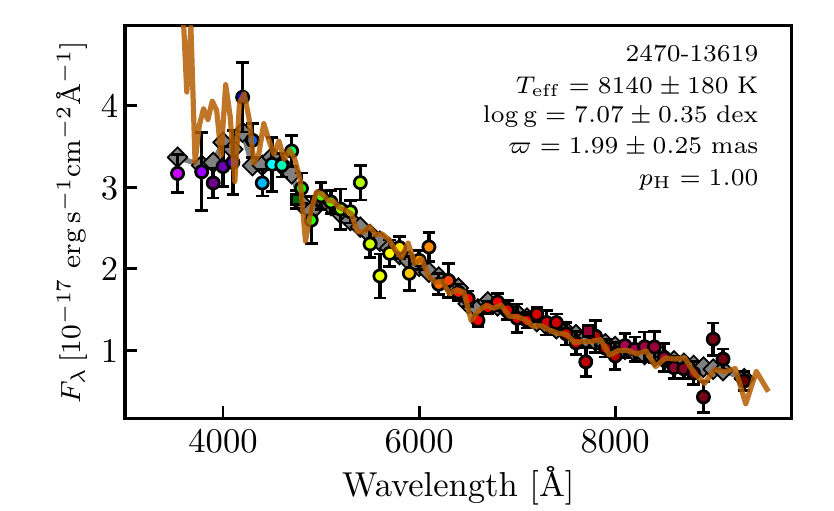}}
\resizebox{0.49\hsize}{!}{\includegraphics{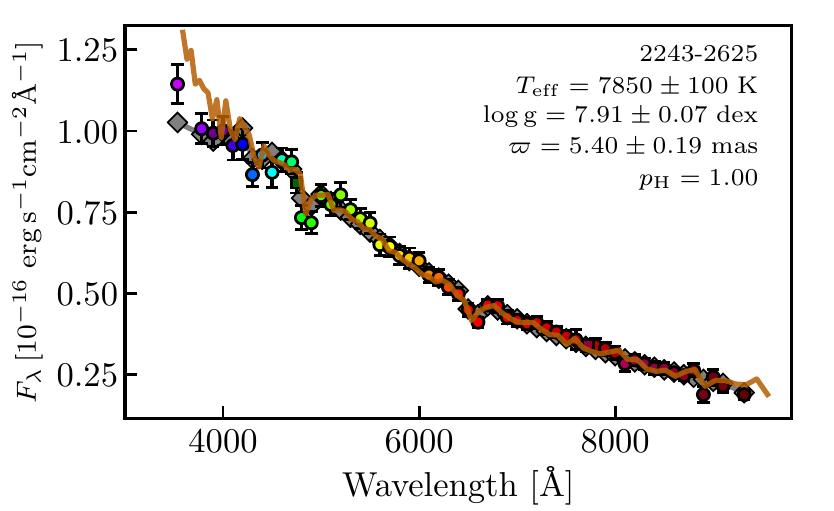}}\\
\resizebox{0.49\hsize}{!}{\includegraphics{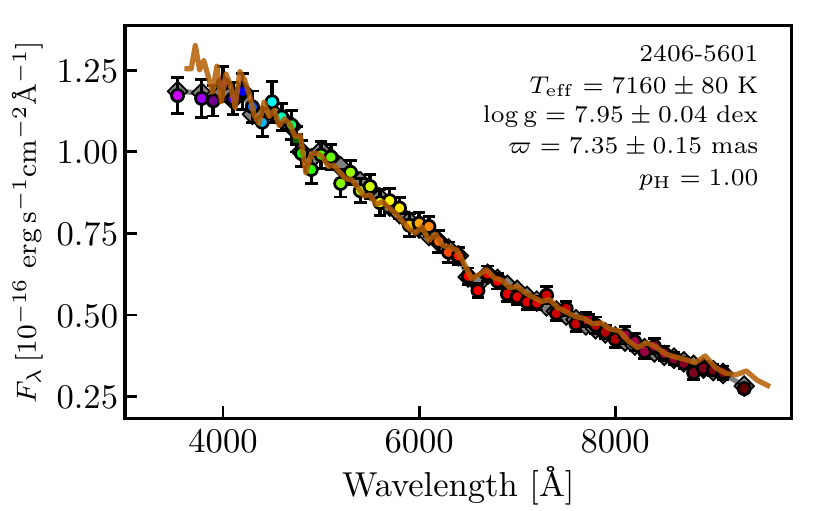}}
\resizebox{0.49\hsize}{!}{\includegraphics{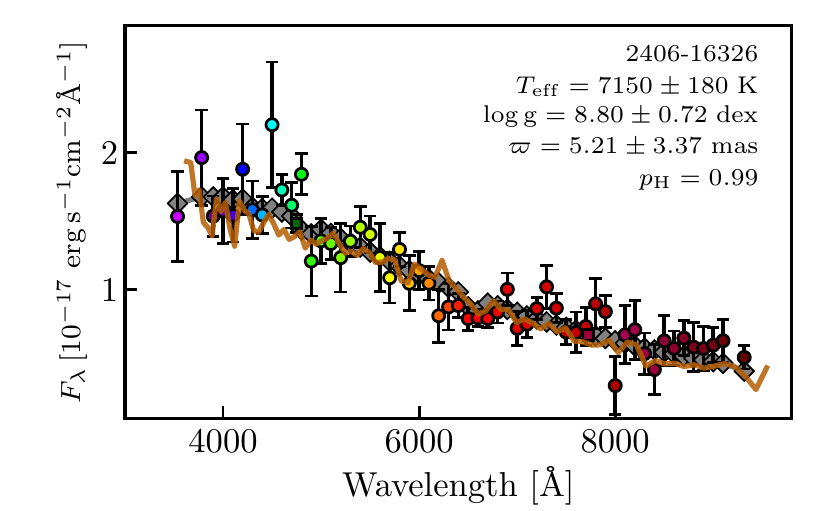}}
\caption{Photo-spectra of the miniJPAS sources classified as H-dominated DAs in descending effective temperature from top-left to bottom-right. Colored circles represent the 56 medium bands, and squares indicate the $g$ and $i$ broad bands. The presented fluxes were estimated from the $3$ arsec diameter aperture photometry corrected for aperture effects (Sect.~\ref{sec:3tototal}) and no correction for interstellar reddening was applied. The gray diamonds show the theoretical flux form the best-fitting model to the data. The parameters of the fitting are labelled in the panel. The brown solid line depicts the SDSS spectra of the sources with a downgraded resolution of $R \sim 90$ for a better comparison. The flux of the SDSS spectra were scaled to match the miniJPAS $r$-band photometry. The flux scale of the SDSS spectra for the sources $2243-2625$, $2406-5601$, and $2406-16326$ has an additional factor $(\lambda/\lambda_0)^{a}$ applied, with $\lambda_0 = 6\,254\,\AA$ and $a = 1.1, 0.3$, and $-0.4$, respectively.}
\label{fig:dah}
\end{figure*}

\begin{figure}[t]
\centering
\resizebox{\hsize}{!}{\includegraphics{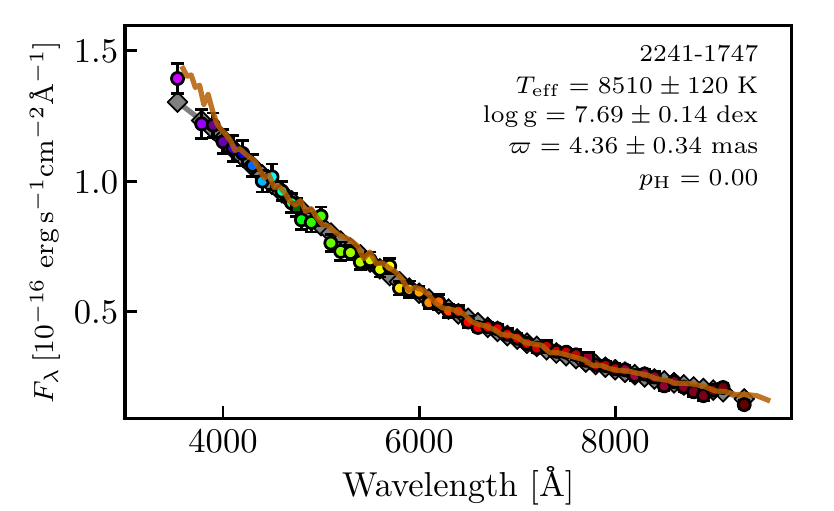}}
\caption{Photo-spectrum of the miniJPAS source 2241-1747 (He-rich DA). Symbols as in Fig. ~\ref{fig:dah}.}
\label{fig:2241_1747}
\end{figure}

The prior probability in the parallax was
\begin{equation}
    P\,(\varpi) = P_{\rm G}\,(\varpi\,|\,\varpi_{\rm EDR3}, \sigma_{\varpi}),
\end{equation}
where $\varpi_{\rm  EDR3}$ and $\sigma_{\varpi}$ are the parallax and its error from ${\it Gaia}$ EDR3 \citep{gaiaedr3,lindegren21a}. The published values of the parallax were corrected using the prescription in \citet{lindegren21b}. In all cases, only positive values of the parallax ($\varpi~>~0$) were allowed.

Finally, the probability of having a H-dominated atmosphere was
\begin{equation}
    p_{\rm H} = \int {\rm PDF}\,({\rm H},\theta)\,{\rm d}\theta.
\end{equation}
The reported values of each parameter in the Tables were estimated by marginalizing over the other parameters at the dominant atmospheric composition defined by $p_{\rm H}$ and performing a Gaussian fit to the obtained distribution. The parameter and its uncertainty are the median and the dispersion of the best-fitting Gaussian. 

\begin{figure}[t]
\centering
\resizebox{\hsize}{!}{\includegraphics{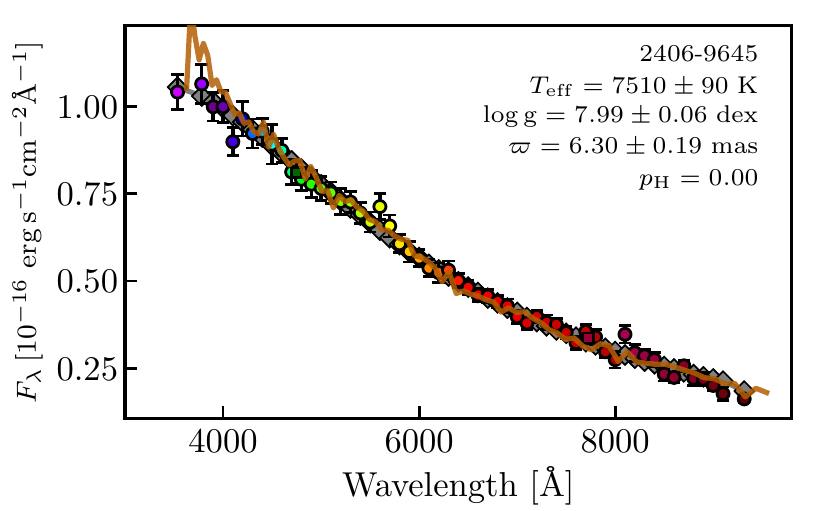}}
\caption{Photo-spectrum of the miniJPAS source 2406-9645 (DC). Symbols as in Fig.~\ref{fig:dah}. The flux scale of the SDSS spectrum has an additional factor $(\lambda/\lambda_0)^{0.3}$ applied, with $\lambda_0 = 6\,254$ \AA.}
\label{fig:2406_9645}
\end{figure}


\section{Analysis of the white dwarf population in miniJPAS}\label{sec:wds}
This section is devoted to the analysis of the white dwarf population in the BPS sample. We provide the relevant individual results for the 11 white dwarfs in Sect.~\ref{sec:onebyone}. The performance in the estimation of the effective temperature and the surface gravity is presented in Sect.~\ref{sec:tefflogg}. The capabilities of the J-PAS filter system to derive the white dwarf atmospheric composition are discussed in Sect.~\ref{sec:spectype}.

\subsection{Notes on individual objects}\label{sec:onebyone}
In this section, we present the relevant results for the 9 DAs (Sect.~\ref{sec:da}), the DC (Sect.~\ref{sec:dc}), and the DZ (Sect.~\ref{sec:dz}) in the BPS sample. All the sources have SDSS spectrum, but in several cases a mismatch between the spectrum and the miniJPAS photometry was evident. Such discrepancies have been also reported by \citet{hollands17}. We found that both data sets can be reconciled by simply multiplying the SDSS spectrum by a factor $(\lambda/\lambda_0)^{a}$, with $\lambda_0 = 6\,254\,\AA$ and $a$ depending on each individual source. 

\subsubsection{DA spectral type}\label{sec:da}
There are eight H-dominated DAs and one He-rich DA in the analyzed sample. These sources are presented in Fig.~\ref{fig:dah}, ordered by decreasing effective temperature. We find that the miniJPAS photometry shows H$\alpha$, H$\beta$, H$\gamma$, and H$\delta$ in most of the cases. The intensity of the Balmer lines is also well recovered by the miniJPAS photo-spectra. We obtained $p_{\rm H} \geq 0.99$ for all the H-dominated DAs. The effective temperature and surface gravity from miniJPAS photometry are compatible with the spectroscopic values at $2\sigma$ level in all the cases (Sect.~\ref{sec:tefflogg}).

The source $2241-1747$ is spectroscopically classified as an He-rich DA by \citet{kepler16}, but was classified as DC in previous studies because of its weak Balmer lines \citep{eisenstein06_dr4,kleinman13}. The analysis of the spectrum with pure-H models implies $T_{\rm eff} \sim 5\,200$ K, but the continuum suggests a hotter system. Both results can be reconciled with a He-dominated atmosphere (see \citealt{rolland18} and \citealt{kilic20}). The miniJPAS data provide a featureless photo-spectrum (Fig.~\ref{fig:2241_1747}) with $p_{\rm H} = 0$ and a shape compatible with the SDSS spectrum of the source. As expected, the photometric effective temperature is $T_{\rm eff} = 8\,510 \pm 120$ K, hotter by $\sim 3\,000$ K than the reported spectroscopic value when a pure-H atmosphere is assumed.

\begin{figure*}[]
\centering
\resizebox{0.49\hsize}{!}{\includegraphics{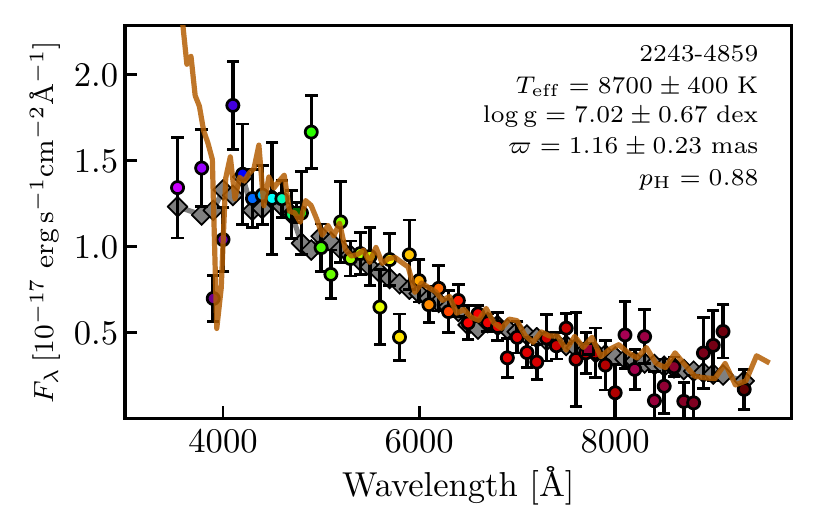}}
\resizebox{0.49\hsize}{!}{\includegraphics{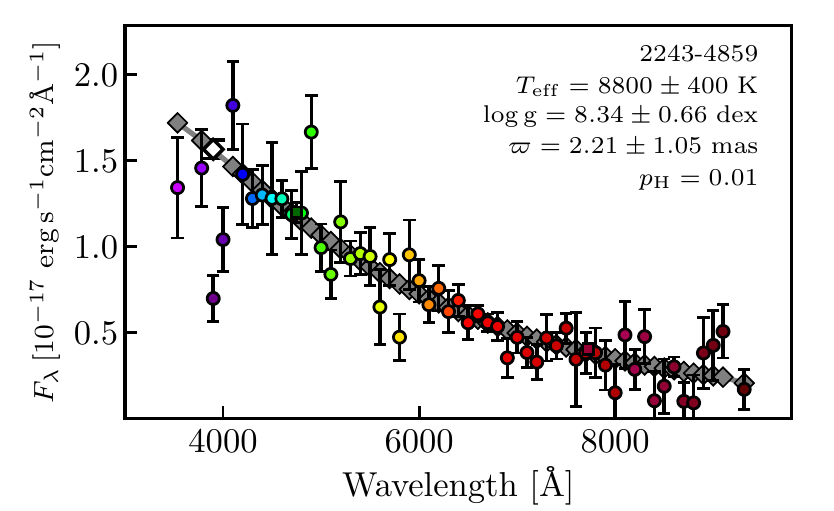}}
\caption{Photo-spectrum of the miniJPAS source 2243-4859 (DZ, $r = 21.4$ mag). Symbols as in Fig.~\ref{fig:dah}. {\it Left panel}: All the miniJPAS passbands were used in the fitting process. {\it Right panel}: The filters $J0390$ and $J0400$ were not included in the fitting. The expected flux in the $J0390$ passband from the modelling is marked by a white diamond.}
\label{fig:2243_4859}
\end{figure*} 

\subsubsection{DC spectral type}\label{sec:dc}
The source $2406-9645$ is the only object in the sample classified as DC (Fig.~\ref{fig:2406_9645}). The miniJPAS photometry is compatible with a featureless continuum, providing $p_{\rm H} = 0$. We estimated $T_{\rm eff} = 7\,510 \pm 90$ K and $\log {\rm g} = 7.99 \pm 0.06$ dex.
    
\subsubsection{DZ spectral type}\label{sec:dz}
The source $2243-4859$ is classified as DZ (calcium white dwarf; Fig.~\ref{fig:2243_4859}), and the \ion{Ca}{ii} H+K absorption feature is present at $3\,950\,\AA$ in the SDSS spectrum. The miniJPAS photometry presents a clear absorption in the passbands $J0390$ and $J0400$. The parameters obtained with all the photometric data provides $p_{\rm H} = 0.88$ and a low surface gravity of $\log {\rm g} = 7.0 \pm 0.7$ dex. We repeated the analysis without the $J0390$ and $J0400$ passbands. The solutions in this case are different, with $p_{\rm H} = 0.01$ and $\log {\rm g} = 8.3 \pm 0.6$ dex. In both cases, the effective temperature is similar, $T_{\rm eff} \sim 8\,800$ K. We note that this object has no parallax information from {\it Gaia} EDR3. We compared the expected flux in the $J0390$ passband from the latter fitting process with the miniJPAS measurement, obtaining an equivalent width of ${\rm EW}_{J0390} = 78 \pm 12\ \AA$, or a 6$\sigma$ detection of the calcium absorption. The J-PAS capabilities to detect metal-polluted white dwarfs are discussed in Sect.~\ref{sec:spectype}.

\subsection{Temperature and surface gravity}\label{sec:tefflogg}
In this section, we compare the $T_{\rm eff}$ and $\log {\rm g}$ values obtained from miniJPAS photometry against those obtained from SDSS spectroscopy by \citet{kepler16, kepler19}, as summarised in Table~\ref{tab:wds_spec}. The DAs spectra were fitted with pure-H models \citep{koester10}, including the Stark-line broadening from \citet{tremblay09} and the 3D corrections from \citet{tremblay13} at $T_{\rm eff} \leq 14\,000$ K. We restricted the comparison to the eight H-dominated white dwarfs in the sample (Sect.\ref{sec:da}) for which the spectroscopic method based on pure-H theoretical models is reliable.

\begin{figure}[t]
\centering
\resizebox{\hsize}{!}{\includegraphics{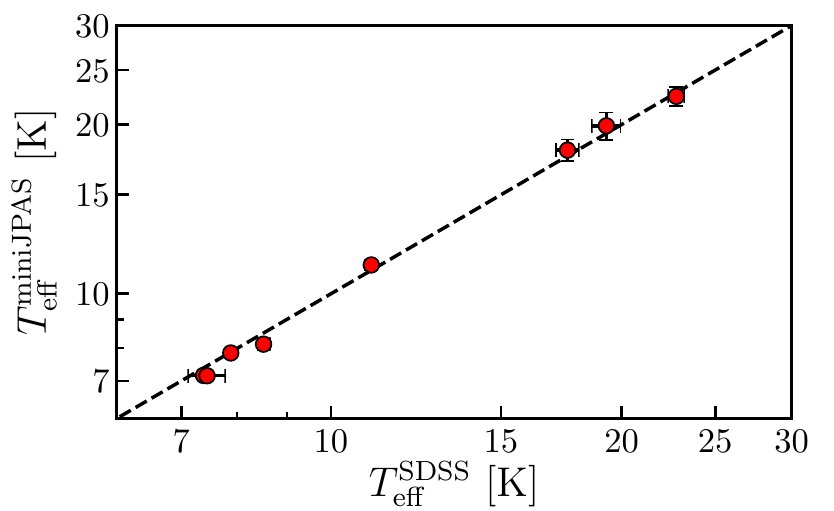}}
\caption{Effective temperature derived from miniJPAS photometry, $T_{\rm eff}^{\rm miniJPAS}$, as a function of the effective temperature derived from SDSS spectrum, $T_{\rm eff}^{\rm SDSS}$ (red dots with black error bars). The dashed line marks the one-to-one relation.} 
\label{fig:teff_sdss}
\end{figure}

We found a tight one-to-one agreement in $T_{\rm eff}$, as illustrated in Fig.~\ref{fig:teff_sdss}. The relative difference between both measurements is $1$\%, with a dispersion of only $3$\%. All the miniJPAS measurements are compatible with the spectroscopic values at $2\sigma$ level. The typical relative error in the effective temperature from miniJPAS data is $2$\%, close to the $1$\% estimated from spectroscopy. Additionally, the typical relative error for the general white dwarf population is $10$\% from the {\it Gaia} EDR3 photometry \citep{GF21} and $5$\% from the J-PLUS photometry \citep{clsj22pda}.

As reported in Sect.~\ref{sec:onebyone}, some SDSS spectra present a shape discrepancy with the miniJPAS photometry. The excellent agreement between the effective temperature from SDSS spectrum, based on the absorption features and thus insensitive to the flux normalization, and from miniJPAS photometry, mainly based on the continuum shape, points to a problematic flux calibration of the discrepant SDSS spectra.

\begin{figure}[t]
\centering
\resizebox{\hsize}{!}{\includegraphics{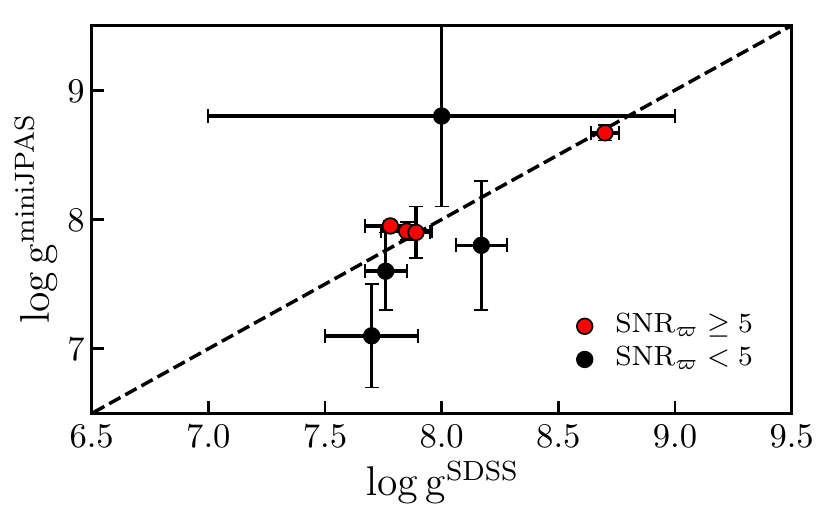}}
\caption{Surface gravity derived from miniJPAS photometry, $\log {\rm g}^{\rm miniJPAS}$, as a function of the surface gravity derived from SDSS spectrum, $\log {\rm g}^{\rm SDSS}$. Red dots mark sources with a ${\rm S/N}$ in $\varpi_{\rm EDR3}$ larger than five, and black dots are sources with a ${\rm S/N}$ lower than five. The dashed line indicates the one-to-one relation.} 
\label{fig:logg_sdss}
\end{figure}

The surface gravity values obtained with both photo-spectra and spectroscopic data are compared in Fig.~\ref{fig:logg_sdss}. We found an agreement between photometric and spectroscopic measurements. However, a precise estimation of $\log {\rm g}$ from miniJPAS photometry demands a precise parallax measurement from {\it Gaia}. The surface gravity information is mainly encoded in the widths of the lines, which are not accessible with the low-resolution miniJPAS photo-spectrum. The assumption of a mass-radius relation in the theoretical models couples the surface gravity and the parallax, so a precise parallax prior from {\it Gaia} astrometry permits to derive the surface gravity when both the effective temperature and the atmospheric composition are well constrained.

We conclude that J-PAS is able to provide effective temperatures with $\sim2$\% precision, but its spectral resolution is not large enough to retrieve a precise surface gravity without parallax information.

\subsection{White dwarf atmospheric composition}\label{sec:spectype}
The main advantage of low-resolution spectral information with respect to broad band photometry is its capability to disentangle the white dwarf main atmospheric composition. The miniJPAS medium-band photometry permits to correctly classify with $99$\% confidence the $11$ white dwarfs in the sample. The hydrogen Balmer lines are visible in miniJPAS photometry of H-dominated atmospheres with temperatures ranging from $7\,000$ K to $22\,000$ K, enabling a clear separation between H- and He-dominated white dwarfs down to $T_{\rm eff} \sim 7\,000$ K. The J-PAS performance at lower and higher effective temperatures will be tested in the near future when larger samples are available.

\begin{figure}[t]
\centering
\resizebox{\hsize}{!}{\includegraphics{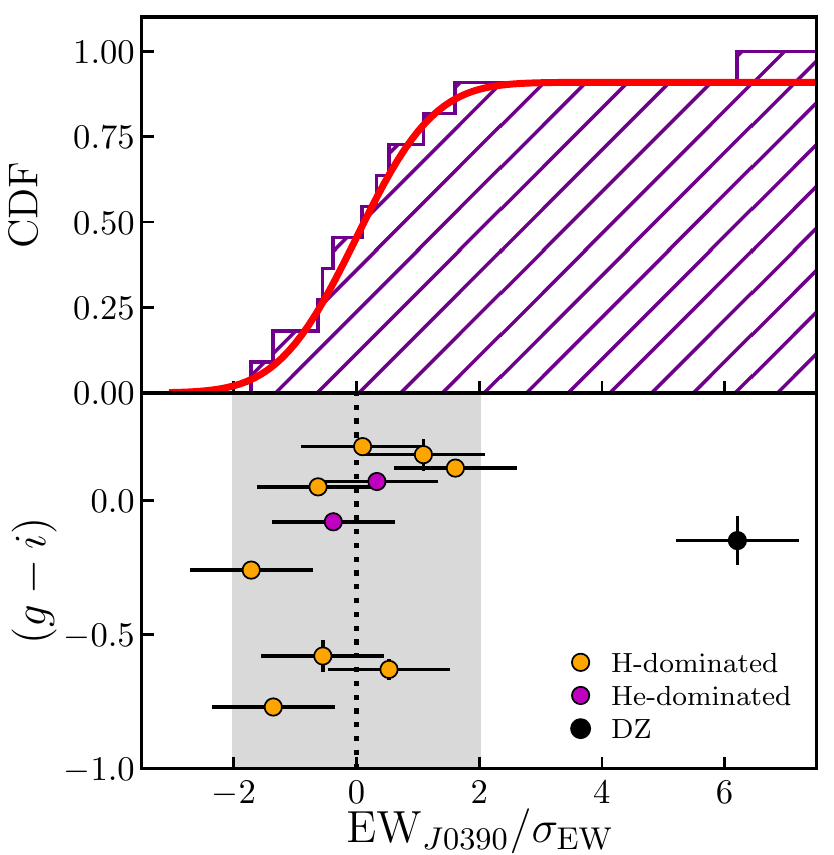}}
\caption{Distribution in the significance of the ${\rm EW}_{J0390}$ measurements as a proxy for calcium absorption in white dwarfs. {\it Top panel}: Cumulative distribution function (CDF) of the ${\rm EW}_{J0390}$ significance. The red solid line is the CDF of a normal distribution normalized to the non-DZ population. {\it Bottom panel}: The color $(g-i)$ as a function of the ${\rm EW}_{J0390}$ significance. Symbols as in Fig.~\ref{fig:bps}. The dotted line marks zero and the gray area shows the $\pm 2\sigma$ interval.}
\label{fig:snew}
\end{figure}

In addition, the presence of polluting metals in the white dwarf atmosphere can be identified thanks to the filters $J0390$ and $J0400$. These passbands are sensitive to the presence of the \ion{Ca}{ii} H+K absorption, as illustrated for the source $2243-4859$ in Fig.~\ref{fig:2243_4859}. We have estimated the equivalent width in the $J0390$ filter as described in Sect.~\ref{sec:dz} for all the white dwarfs in the sample. The significance of the measurement, estimated as ${\rm EW}_{J0390}/\sigma_{\rm EW}$, is presented in Fig.~\ref{fig:snew}. Those sources classified as non-DZ cluster around zero and are compatible with the absence of calcium absorption at $2\sigma$ level. A Kolmogov-Smirnov test provides a $98$\% probability that their distribution is drawn for a normal distribution, as expected if the measurements are compatible with zero within uncertainties. The only outlier is the DZ source, which presents a $6\sigma$ detection. Thus, the ${\rm EW}_{J0390}$ measurement can be used to select new metal-polluted white dwarfs. In addition to the calcium absorption, the presence of other prominent absorption features in cool white dwarfs, such as the \ion{Mg}{i} $b$ triplet and the \ion{Na}{i} doublet at $5\,893\ \AA$ (e.g., \citealt{hollands17}), would be also detectable in the J-PAS data.

Our results demonstrate that J-PAS photometry will be able to provide the main atmospheric composition for white dwarfs, including the presence of polluting metals. However, the small sample available in the miniJPAS area does not permit to study in detail the J-PAS capabilities in the estimation of either H-to-He abundances on hybrid types (e.g., DABs and DBAs) or metal abundances in polluted systems, and does not contain magnetic, carbon, or peculiar white dwarfs. The performance of the J-PAS filter system with these types will be evaluated in the future when larger samples are observed. 

We conclude that J-PAS photo-spectrum will allow to accurately study the evolution of He-dominated white dwarfs and the fraction of metal-polluted white dwarfs with effective temperature at least down to $T_{\rm eff} \sim 7\,000$ K.

\section{White dwarf selection based on miniJPAS photometry}\label{sec:qso}
We have demonstrated the capabilities of multi-band photometry in the study of known white dwarfs. This will impact the analysis of the future white dwarf samples, like those expected from {\it Gaia} data. In addition, we aim to test the performance of miniJPAS data to select new white dwarf candidates just based on optical photo-spectra. In this section, we analysed the BPS sample on this regard. For that, the model flux presented in Sect. 3 was simplified as:
\begin{equation}
    f_{t,j}^{\rm mod}\,(t,T_{\rm eff},\log g) = C_{r}\,F_{t,k}\,(T_{\rm eff},\log g)\,10^{0.4\,k_j\,E(B-V)}\,10^{0.4\,C^{\rm aper}_j},
\end{equation}
where $C_{r}$ is a constant to normalise the theoretical flux to the measured flux in the miniJPAS $r$ band. That is, we assumed a unique scale for each \{$T_{\rm eff}$, $\log {\rm g}$\} pair to remove the parallax as a parameter in the fitting process. The prior in parallax was also neglected to provide a consistent analysis of Galactic and extragalactic sources, and only the likelihood of being white dwarf was computed. The assumed colour excess was computed at the distance implied by the $C_
r$ normalization.

We used this scheme to obtain the minimum $\chi^2$ for each object as
\begin{equation}
    \chi^2_{\rm WD} = -2 \times \log \mathcal{L}_{\rm max},
\end{equation}
where $\mathcal{L}_{\rm max}$ is the maximum likelihood obtained in the exploration of the parameters space for both H- and He-dominated atmospheres.

\begin{figure}[t]
\centering
\resizebox{\hsize}{!}{\includegraphics{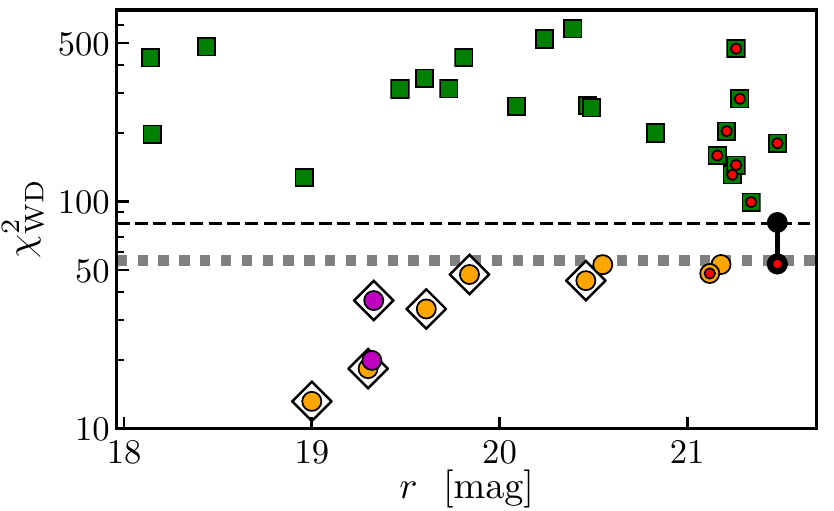}}
\caption{Minimum $\chi^2_{\rm WD}$ as a function of the $r$-band magnitude for the BPS. Circles mark spectroscopic white dwarfs (H-dominated: orange, He-dominated: purple, metal polluted: black), and green squares show QSOs. The values obtained with and without the filters $J0390$ and $J0410$ for the calcium white dwarf $2243-4859$ are connected by a black line. Those sources without {\it Gaia} EDR3 parallax information are marked with a red dot. Those white dwarfs included in the \citet{GF21} catalog are marked with a white diamond. The black dashed line depicts the separation between white dwarfs and QSOs, $\chi^2_{\rm WD} = 80$. The dotted line shows the expected value for white dwarfs given the degrees of freedom in the analysis, $\chi^2_{\rm WD} = 55$.}
\label{fig:rml}
\end{figure}

We present the results in Fig.~\ref{fig:rml}, and show the correspoding $\chi^2_{\rm WD}$ in Tables~\ref{tab:wds_jpas} and \ref{tab:qso}. We found a clear separation between white dwarfs and QSOs in the BPS sample in terms of their $\chi_{\rm WD}^2$, with white dwarfs having lower values.

We have $58$ photometric points and three effective parameters when the constraints from the {\it Gaia} EDR3 parallax are weak \citep{clsj22pda}. Thus, the values should tend to $\chi^{2}_{\rm WD} \approx 55$. The white dwarfs tend to $\chi^{2}_{\rm WD} \approx 53$ at the faint end, as expected. There is also a trend towards lower $\chi_{\rm WD}^2$ at brighter magnitudes ($r \lesssim 19.5$ mag), reflecting an overestimation of the uncertainty in the photometric calibration, that was set to $\sigma_{\rm cal} = 0.04$ mag for all the passbands. 
The QSOs have larger values of $\chi_{\rm WD}^2$, reaching even $\chi_{\rm WD}^2 = 500$. This is due to the presence of emission lines in the photo-spectrum, clearly off from the expected absorption lines or the featureless spectrum for white dwarfs. The presence of the Lyman $\alpha$ line in the QSO spectrum at $z > 2$ provides the most prominent differences.

We conclude that white dwarfs in the BPS sample can be selected with high confidence by imposing $\chi_{\rm WD}^2 \leq 80$. High-purity white dwarf samples will be defined with J-PAS, complementing the astrometric information from ${\rm Gaia}$ down to $G \sim 21$ and permitting to extend the analysis beyond {\it Gaia} capabilities. As an example, of the $33$ sources in the BPS sample, $8$ (25\%) do not have an entry in the {\it Gaia} EDR3 catalog and only two of them are white dwarfs (Fig.~\ref{fig:rml}).

The calcium white dwarf $2243-4859$ presents $\chi_{\rm WD}^2 = 81$ if all the passbands are used in the fitting, a value that decreases to $\chi_{\rm WD}^2 = 53$ when $J0390$ and $J0400$ are removed from the analysis. This implies that these two filters are clearly discrepant with the expected white dwarf flux due to the presence of calcium absorption, and provide a way to select metal-polluted white dwarfs using multi-filter photometry (Sect.~\ref{sec:spectype}). We checked that no QSO were located below the $\chi_{\rm WD}^2 = 80$ limit when the $J0390$ and $J0400$ passbands are removed from the analysis.

Finally, we searched for white dwarf candidates in the {\it Gaia}-based catalog presented by \citet{GF21}. Following the authors' suggestion, we only kept those sources with a white dwarf probability larger than $0.75$. We found six sources with $r < 20.5$ mag (Fig.~\ref{fig:rml} and Table~\ref{tab:wds}). Two of the four sources with $r > 20.5$ mag present low ${\rm S/N}$ in the parallax and the other two have no parallax measurement. There is one bright source not included in the catalog, $2241-1747$. This source was discarded by \citet{GF21} because of the presence of a fainter, close source that increases the number of parameters in the solved astrometric solution\footnote{The values of the parameters \texttt{ASTROMETRIC\_PARAMS\_SOLVED} $= 95$, \texttt{ASTROMETRIC\_EXCESS\_NOISE} $= 2.2$, and \texttt{ASTROMETRIC\_EXCESS\_NOISE\_SIG} $= 8.3$ do not fulfill the requirements imposed by Eq.~(8) in \citet{GF21}}. This exercise suggests that we could double the number of high-confidence white dwarfs in the J-PAS area with respect to {\it Gaia}-based catalogs.

A complete analysis of the BPS sample demands the addition of QSO models. This is beyond the scope of the present paper, and we demonstrated that the comparison between miniJPAS photometry and the white dwarf theoretical models is enough to discriminate the QSOs in the bluest sources at $r \leq 21.5$ mag thanks to the $56$ medium bands in the J-PAS photometric system.

\section{Discussion and conclusions}\label{sec:conclusion}
We have analysed the physical properties of 11 white dwarfs in the miniJPAS data set, which provides low-resolution photo-spectrum thanks to a unique filter system of $56$ medium bands with ${\rm FWHM} \approx 145\ \AA$ covering continuously the optical range from $3\,500$ to $9\,300$ \AA.

We found that the effective temperature determination has a typical relative error of $2$\%, whereas the estimation of a precise surface gravity demands the parallax information from {\it Gaia}. Regarding the atmospheric composition, the J-PAS filter system is able to correctly classify H- and He-dominated atmosphere white dwarfs at least in the temperature range covered by the miniJPAS white dwarf sample, $7\,000 < T_{\rm eff} < 22\,000$ K. We also show that the presence of polluting metals can be revealed by the \ion{Ca}{ii} H+K absorption, as traced by the $J0390$ and $J0400$ passbands. Furthermore, the miniJPAS low-resolution information is able to disentangle between white dwarfs with $T_{\rm eff} \gtrsim 7\,000$ K and extragalactic QSOs with similar broad-band colors.

The J-PAS project, which will survey thousands of square degrees in the northern sky, will provide a unique data set of several tens thousand white dwarfs down to $r \sim 21.5$ mag to analyse the fraction of He-dominated white dwarfs with $T_{\rm eff}$, search for new metal-polluted systems, derive the white dwarf luminosity function, detect unusual objects, etc.

In addition to the data-driven forecast for J-PAS, our results provide hints about the performance of the future {\it Gaia} DR3 spectro-photometry, complementing the theoretical expectations presented in \citet{carrasco14_wd}. The {\it Gaia} spectro-photometry will have a comparable resolution to miniJPAS data, $R = 30-90$, and therefore similar capabilities are expected at the same ${\rm S/N}$ level. 

There are also relevant synergies between {\it Gaia} and J-PAS that are worth noticing. On the one hand, {\it Gaia} provides a full sky data set. On the other hand, J-PAS is deeper and provides high ${\rm S/N}$ photo-spectrum even at $G = 21$ mag. We envision three different regimes: (i) Bright sources with enough ${\rm S/N}$ in {\it Gaia} spectro-photometry. Two independent measurements of the white dwarf properties will be available, providing insights about systematic errors in both surveys and tests for the {\it Gaia} capabilities at the lower ${\rm S/N}$. (ii) Faint sources with enough ${\rm S/N}$ in {\it Gaia} astrometry. The combination of J-PAS photo-spectra and {\it Gaia} parallaxes will permit to define and study in detail the white dwarf population down to $G \sim 21$ mag. And (iii) white dwarf candidates beyond {\it Gaia} data, $G > 21$ mag. The J-PAS photo-spectrum can provide clean samples of white dwarfs for spectroscopic follow-up in a magnitude range dominated by QSOs and without the parallax information from {\it Gaia}. In this range, the main alternative will be the use of reduced proper motions for deep, multi-epoch surveys such as the Legacy Survey for Space and Time (LSST, \citealt{lsst}), capable to obtain reliable white dwarf candidates down to $G \sim 23$ mag \citep{fantin20}.

As an illustrative example, the $11$ white dwarfs in the miniJPAS area can be split as: six sources ($55$\%) included in the \citet{GF21} catalog based on {\it Gaia} EDR3 with a white dwarf probability larger than $0.75$, three sources ($27$\%) with low ${\rm S/N}$ or a low quality flag in {\it Gaia} and not included in the \citet{GF21} catalog, and two white dwarfs ($18$\%) without an entry on the {\it Gaia} EDR3 catalog. Hence, there is potential to double the number of high-confidence white dwarf candidates in the future J-PAS area with respect to the {\it Gaia}-based catalogs. However, this will depend on the ${\rm S/N}$ achieved at the final {\it Gaia} data release.

Finally, the J-PAS photo-spectra will complement the spectroscopic follow-up of the {\it Gaia}-selected white dwarf population planned with SDSS-V, WEAVE, and DESI in the northern sky. J-PAS will detect and characterize new white dwarfs beyond the {\it Gaia} limits, improving the selection function of the spectroscopic surveys and providing extra candidates for spectroscopic follow up.

\begin{acknowledgements}
We dedicate this paper to the memory of our six IAC colleagues and friends who met with a fatal accident in Piedra de los Cochinos, Tenerife, in February 2007, with  special thanks to Maurizio Panniello, whose teachings of \texttt{python} were so important for this paper.

This paper has gone through internal review by the J-PAS collaboration, with relevant comments and suggestions from A.~Bragaglia and A.~Alvarez-Candal.


Based on observations made with the JST250 telescope and PathFinder camera for the miniJPAS project at the Observatorio Astrof\'{\i}sico de Javalambre (OAJ), in Teruel, owned, managed, and operated by the Centro de Estudios de F\'{\i}sica del Cosmos de Arag\'on (CEFCA). We acknowledge the OAJ Data Processing and Archiving Unit (UPAD) for reducing and calibrating the OAJ data used in this work. Funding for OAJ, UPAD, and CEFCA has been provided by the Governments of Spain and Arag\'on through the Fondo de Inversiones de Teruel; the Aragonese Government through the Research Groups E96, E103, E16\_17R, and E16\_20R; the Spanish Ministry of Science, Innovation and Universities (MCIU/AEI/FEDER, UE) with grant PGC2018-097585-B-C21; the Spanish Ministry of Economy and Competitiveness (MINECO/FEDER, UE) under AYA2015-66211-C2-1-P, AYA2015-66211-C2-2, AYA2012-30789, and ICTS-2009-14; and European FEDER funding (FCDD10-4E-867, FCDD13-4E-2685). Funding for the J-PAS Project has been provided by the Governments of Spain and Arag\'on through the Fondo de Inversiones de Teruel, European FEDER funding and the Spanish Ministry of Science, Innovation and Universities, and by the Brazilian agencies FDNCT, FINEP, FAPESP, FAPERJ and by the National Observatory of Brazil. Additional funding was also provided by the Tartu Observatory and by the J-PAS Chinese Astronomical Consortium.

P.~-E.~T. has received funding from the European Research Council under the European Union's Horizon 2020 research and innovation programme n. 677706 (WD3D). 

A.~E. and J.~A.~F.~O. acknowledge the financial support from the Spanish Ministry of Science and Innovation and the European Union - NextGenerationEU through the Recovery and Resilience Facility project ICTS-MRR-2021-03-CEFCA.

M.~A.~G. is funded by the Spanish Ministerio de Ciencia, Innovaci\'on y Universidades (MCIU) grant PGC2018-102184-B-I00, co-funded by FEDER funds. He also acknowledges support from the State Agency for Research of the Spanish MCIU through the ‘Center of Excellence Severo Ochoa’ award to the Instituto de Astrof\'\i sica de Andaluc\'\i a (SEV-2017-0709).

J.~V. acknowledges the technical members of the UPAD for their invaluable work: Juan Castillo, Tamara Civera, Javier Hern\'andez, \'Angel L\'opez, Alberto Moreno, and David Muniesa.

F.~M.~J.~E. acknowledges financial support from the Spanish MINECO/FEDER through the grant AYA2017-84089 and MDM-2017-0737 at Centro de Astrobiología (CSIC-INTA), Unidad de Excelencia María de Maeztu, and from the European Union’s Horizon 2020 research and innovation programme under Grant Agreement no. 824064 through the ESCAPE - The European Science Cluster of Astronomy \& Particle Physics ESFRI Research Infrastructures project.

R.~L.~O. acknowledges financial support from the Brazilian institutions CNPq (PQ-312705/2020-4) and FAPESP (\#2020/00457-4).

R.~A.~D. acknowledges support from the Conselho Nacional de Desenvolvimento Cient\'{\i}fico e Tecnol\'ogico - CNPq through BP grant 308105/2018-4, and the Financiadora de Estudos e Projetos - FINEP grants REF. 1217/13 - 01.13.0279.00 and REF 0859/10 - 01.10.0663.00 and also FAPERJ PRONEX grant E-26/110.566/2010 for hardware funding support for the J-PAS project through the National Observatory of Brazil and Centro Brasileiro de Pesquisas F\'{\i}sicas.

L.~S.~J. acknowledges the support of CNPq (304819/2017-4) and FAPESP (2019/10923-5).

This work has made use of data from the European Space Agency (ESA) mission
{\it Gaia} (\url{https://www.cosmos.esa.int/gaia}), processed by the {\it Gaia} Data Processing and Analysis Consortium (DPAC, \url{https://www.cosmos.esa.int/web/gaia/dpac/consortium}). Funding for the DPAC has been provided by national institutions, in particular the institutions participating in the {\it Gaia} Multilateral Agreement.

Funding for the Sloan Digital Sky Survey IV has been provided by the Alfred P. Sloan Foundation, the U.S. Department of Energy Office of Science, and the Participating  Institutions. SDSS-IV acknowledges support and resources from the Center for High Performance Computing  at the  University of Utah. The SDSS  website is \url{www.sdss.org}. SDSS-IV is managed by the Astrophysical Research Consortium for the Participating Institutions of the SDSS Collaboration including the Brazilian Participation Group, the Carnegie Institution for Science, Carnegie Mellon University, Center for Astrophysics | Harvard \& Smithsonian, the Chilean Participation Group, the French Participation Group, Instituto de Astrof\'isica de Canarias, The Johns Hopkins University, Kavli Institute for the Physics and Mathematics of the Universe (IPMU) / University of Tokyo, the Korean Participation Group, Lawrence Berkeley National Laboratory, Leibniz Institut f\"ur Astrophysik Potsdam (AIP),  Max-Planck-Institut f\"ur Astronomie (MPIA Heidelberg), Max-Planck-Institut f\"ur Astrophysik (MPA Garching), Max-Planck-Institut f\"ur Extraterrestrische Physik (MPE), National Astronomical Observatories of China, New Mexico State University, New York University, University of Notre Dame, Observat\'ario Nacional / MCTI, The Ohio State University, Pennsylvania State University, Shanghai Astronomical Observatory, United Kingdom Participation Group, Universidad Nacional Aut\'onoma de M\'exico, University of Arizona, University of Colorado Boulder, University of Oxford, University of Portsmouth, University of Utah, University of Virginia, University of Washington, University of Wisconsin, Vanderbilt University, and Yale University.

This research has made use of the SIMBAD database, operated at CDS, Strasbourg, France.

This research made use of \texttt{Astropy}, a community-developed core \texttt{Python} package for Astronomy \citep{astropy}, and \texttt{Matplotlib}, a 2D graphics package used for \texttt{Python} for publication-quality image generation across user interfaces and operating systems \citep{pylab}.
\end{acknowledgements}

\bibliographystyle{aa}
\bibliography{biblio}

\end{document}